\documentclass[prl,twocolumn,showpacs]{revtex4}
\usepackage[dvips]{graphicx}
\input{epsf}
\usepackage{amsmath}
\usepackage{amssymb}
\usepackage[mathscr]{eucal}
\usepackage{bm}
\usepackage{theorem}

\newcommand{\beq}{\begin{equation}}
\newcommand{\eeq}{\end{equation}}
\newcommand{\beqa}{\begin{eqnarray}}
\newcommand{\eeqa}{\end{eqnarray}}
\newcommand{\beqan}{\begin{eqnarray*}}
\newcommand{\eeqan}{\end{eqnarray*}}

\newcommand{\tr}[1]{{\rm tr} \left( #1 \right) }
\newcommand{\ket}[1]{| #1 \rangle}
\newcommand{\bra}[1]{\langle #1 |}

\newcommand{\ox}{\otimes}

{\theorembodyfont{\upshape}

}

\begin{document}

  \title{Characterizing 3-qubit UPB states: violations of LHV models, preparation via nonlocal unitaries and PPT entangled nonlocal orbits}
\author{Claudio Altafini}
% \thanks{This work was supported by a grant from the Foundation Blanceflor Boncompagni-Ludovisi.}
\affiliation{SISSA-ISAS  \\
International School for Advanced Studies \\
via Beirut 2-4, 34014 Trieste, Italy }


% 03.65.Ud Entanglement and quantum nonlocality
% 03.67.-a Quantum information
% 03.67.Mn Entanglement production, characterization, and manipulation 
% 03.67.Lx Quantum computation

\begin{abstract}
For the 3-qubit UPB state, i.e., the bound entangled state constructed from an Unextendable Product Basis of Bennett {\em et al.} \cite{Bennett1}, we provide a set of violations of Local Hidden Variable (LHV) models based on the particular type of reflection symmetry encoded in this state.
The explicit nonlocal unitary operation needed to prepare the state from its reflected separable mixture of pure states is given, as well as a nonlocal one-parameter orbit of states with Positive Partial Transpositions (PPT) which swaps the entanglement between a state and its reflection twice during a period.

\end{abstract}

\maketitle 

% \begin{keyword}
% matrix Lie groups \sep time-varying differential equations \sep Magnus expansions \sep Wei-Norman formula.
% \end{keyword}

A bound entangled state is a nonseparable quantum state for which no distillation process is possible \cite{Horodecki7}.
While some multiparty bound entangled states are easier to detect because they show some form of bipartite entanglement through some of the cuts \cite{Smolin1,Dur1}, the bound entangled states built from an Unextendable Product Basis (UPB states) of \cite{Bennett1,DiVincenzo1} are probably the most mysterious one to date.
For them, in fact, all partial transpositions are positive (PPT), no violation of Bell inequality is known and no criterion exist for detecting their entanglement.
They are defined as the ``complement'' of a mixture (with all equal weights) of reciprocally orthogonal product states (such that no other product state orthogonal to all members exists) in the given Hilbert space.
In other words, since any separable (pure or mixed) state is a convex combination of pure product states \cite{Horodecki2}, if we take an orthogonal basis of such states which is unextandable, i.e., such that there does not exist any other product state orthogonal to the basis, then states which belong to the complement of this incomplete space in the Hilbert space of the quantum system are guaranteed to be nonseparable.
If the mapping of the given mixture is done as in \cite{Bennett1} mixing the UPB basis (with a minus sign in front) with the random state, then the property of PPT is also guaranteed.
This mapping is nonunitary because of the minus sign just mentioned and it is shown in \cite{Cla-qu-ent-reflx1} to be a particular case of a class of mirror-like symmetries whose origin and behavior are easily visible once we use the real tensorial parametrization proposed in \cite{Cla-qu-ent1}.
Such a {\em reflection} symmetry is nonequivalent to any known quantum symmetry and makes sense only for mixed states of multiqubits as the UPB states discussed here.
It sheds considerable light into the structure of the bound entanglement of these states, as for instance it allows to identify violations of LHV models involving 3 commuting observables, one violation for each term of the separable mixture that is reflected.
We will see it for the 3-qubit UPB state (which is the quantum state considered in the entire paper).

Two other issues are discussed after that: first the preparation of a UPB state, starting from the corresponding separable mixture and using nonlocal unitary operations; second the existence of nonlocal orbits which are PPT.
We have found a single nontrivial one-parameter such orbit whose peculiarity is that the bound entanglement swaps between a state and its reflection twice during the orbit's period.

Consider the 3-qubit density operator corresponding to the UPB state $ \rho_{\rm UPB} = \frac{1}{4} \left( \openone_8 - \sum_{j=1}^4 \ket{\psi_j} \bra{\psi_j} \right) $, with $ \ket{\psi_j} = \ket{01+}, \ket{1+0},\ket{+01}, \ket{---} $ (where $\ket{\pm} = \frac{1}{\sqrt{2}} \left( \ket{0} \pm \ket{1} \right) $) introduced in \cite{Bennett1}.
In terms of the tensor of coherences \footnote{The formalism of the tensor of coherences \cite{Cla-qu-ent1} and its infinitesimal counterpart \cite{Cla-spin-tens1} will be used throughout the paper without further notice.}, if $ x=\frac{1}{8 \sqrt{2}} $, $ \rho_{\rm UPB} $ is:
\beq
\rho_{\rm UPB} :
\begin{cases}
& \varrho^{000} = \frac{1}{2\sqrt{2} } \\
& \varrho^{\{ 031,\,033,\,103,\,111,\,133,\,303,\,310,\,313,\,330,\,331\} }=x  \\
& \varrho^{\{ 011,\,013,\,101,\,110,\,130,\,301\}}=-x  \\
& \varrho^{jkl} =0 \qquad \text{otherwise} .
\end{cases}
\label{eq:varrho-UPB}
\eeq 
The ``complement'' mentioned above is identified by the following operation: 
\beq 
\rho_{\rm UPB } = \frac{1}{4}\openone_8 - \rho_{\rm sep } ,
\label{eq:mapUPB}
\eeq 
where $ \rho_{\rm sep } =  \frac{1}{4} \sum_{j=1}^4 \ket{\psi_j} \bra{\psi_j } $.
If we call $ \rho_{{\rm sep}_j} = \ket{\psi_j} \bra{\psi_j} $, in terms of the local Bloch vectors, $ \rho_{\rm sep }=  \frac{1}{4} \sum_{j=1}^4 \rho_{{\rm sep}_j}  $ has components ($ \varrho^0=\varrho^1=\varrho^3=\frac{1}{\sqrt{2}}$): 
\beq
\begin{split}
\rho_{{\rm sep}_1} & =  
    \left( \varrho^0 \lambda_0 + \varrho^{3} \lambda_{3} \right) 
\ox \left( \varrho^0 \lambda_0 - \varrho^{3} \lambda_{3} \right) 
\ox \left( \varrho^0 \lambda_0 + \varrho^{1} \lambda_{1} \right) ,
\\
\rho_{{\rm sep}_2} & =  
    \left( \varrho^0 \lambda_0 - \varrho^{3} \lambda_{3} \right) 
\ox \left( \varrho^0 \lambda_0 + \varrho^{1} \lambda_{1} \right) 
\ox \left( \varrho^0 \lambda_0 + \varrho^{3} \lambda_{3} \right) ,
\\
\rho_{{\rm sep}_3} & =  
    \left( \varrho^0 \lambda_0 + \varrho^{1} \lambda_{1} \right) 
\ox \left( \varrho^0 \lambda_0 + \varrho^{3} \lambda_{3} \right) 
\ox \left( \varrho^0 \lambda_0 - \varrho^{3} \lambda_{3} \right) ,
\\
 \rho_{{\rm sep}_4} & =  
    \left( \varrho^0 \lambda_0 - \varrho^{1} \lambda_{1} \right) 
\ox \left( \varrho^0 \lambda_0 - \varrho^{1} \lambda_{1} \right) 
\ox \left( \varrho^0 \lambda_0 - \varrho^{1} \lambda_{1} \right) .
\end{split}
\label{eq:rho_UPBsep}
\eeq 
In \cite{Cla-qu-ent-reflx1}, it is shown that for the tensor of coherences the operation \eqref{eq:mapUPB} is a reflection symmetry applied to the joint density of the 3 qubits, which results in a change of sign to the homogeneous part of the tensor, i.e., to all the expectation values of 1-, 2- and 3-qubit observables but not to the trivial one $ \varrho^{000}= \tr{\rho_{\rm sep} \Lambda_{000} } $ that ``takes care'' of the trace: if $\rho_{\rm sep} = \varrho^{000} \Lambda_{000} + \xi $, then $  \rho_{\rm UPB } = \bar{S}_{64} ( \rho_{\rm sep } ) = \varrho^{000} \Lambda_{000} - \xi $, in the notation of \cite{Cla-qu-ent-reflx1}.
The map $ \bar{S}_{4^3} $ is a nonunitary linear map that preserves trace and Hermiticity.
It is not well-defined in the set of admissible densities of 3 qubits, call it ${\cal D} $, although it is in the smaller subset $  {\cal C} = \left\{ \rho \in {\cal D} \text{ s.t. }  {\rm eig}(\rho) \in [0, \, \frac{1}{4} ] \right\} $ where it is even invariant: $ \rho \in {\cal C} $ implies $ \tilde{\rho} = \bar{S}_{64} (\rho) \in {\cal C} $.
Obviously $ \rho_{\rm sep }\in {\cal C} $ and $ \rho_{\rm UPB}\in {\cal C} $.
Because of the all equal weights, the mapping \eqref{eq:mapUPB} is also isospectral: $  {\rm eig}(\rho_{\rm sep }) =  {\rm eig}(\rho_{\rm UPB}) =\{ 0, \, \frac{1}{4} \} $ of multiplicity 4.
The reflection operation (restricted to $ {\cal C} $) is obviously more general than the UPB construction of \cite{Bennett1} and needs not yield a PPT density. For that the ``unextendability'' of UPB is required.
No one of the 4 components $ \rho_{{\rm sep}_j} $ taken alone (each is obviously a separable density) is a density when reflected.
For example $ {\rm eig} \left( \bar{S}_{64}  \left(\rho_{{\rm sep}_1}  \right) \right) = \{ - \frac{3}{4}, \frac{1}{4} \}  $ of multiplicity resp. 1 and 7.
Hence, although {\em a posteriori}  (i.e., after the cancellations), it holds that $ \bar{S}_{64} \left( \frac{1}{4} \sum_{j=1}^4 \rho_{{\rm sep}_j} \right) =  \frac{1}{4} \sum_{j=1}^4 \bar{S}_{64} \left(  \rho_{{\rm sep}_j} \right)$, ``formally'' the right hand side contains something else than just density operators.
The convex statistical weights are fundamental, thus confirming that operations like  \eqref{eq:mapUPB} are intrinsically defined only for mixed states.
For the same reason the use and the meaning of the ``complement'' operation of \cite{Bennett1} must be handled with care.

Look at the tensor of coherences of the 4 components of $ \rho_{\rm sep} $ as obtained from \eqref{eq:rho_UPBsep}.
It is straightforward to check that in the sum all and only the 1-qubit coherences are canceled.
Such cancellations imply that all 1-qubit reduced densities are random.
A first consequence is that the same mapping $ \rho_{\rm sep} \to\rho_{\rm UPB} $ can be obtained by means of two-qubit partial reflections: $\left( \bar{S}_{16}\otimes \openone_4   \right) (\rho_{\rm sep} ) = \rho_{\rm UPB} $,
and similarly for the other two.
A second, more important, consequence is that 1-spin generators ($ -i {\rm ad}_{\Lambda_{j00}} $ etc.) have no effect on $ \rho_{\rm UPB} $.
Said otherwise, $ \rho_{\rm UPB} $ is invariant to LOCC, even an infinite amount of them, even stochastic, since the random reduced density is a fixed point also in the class of local filtering operations. 
A third consequence is that the 3-qubit bound entanglement cannot be modified by symmetric extensions of tensorial type \cite{Terhal2}, neither by convex combinations of extensions.
From the expression $ \rho_{\rm UPB} = \varrho^{000} \Lambda_{000} - \xi $, it is fairly easy to conclude on the character of any tensor product $\rho_{\rm UPB}\ox \rho_{\rm a} $ of $ \rho_{\rm UPB} $ with an ancilla $ \rho_{\rm a} $.
In fact, if for example the ancilla is a qubit $ \rho_{\rm a} = \varrho^0 \lambda_0 + \varrho^j_{\rm a} \lambda_j $, then $ \rho_{\rm UPB} \ox \rho_{\rm a} = \varrho^{0000} \Lambda_{0000} + \varrho^{000}\varrho^j_{\rm a}  \Lambda_{000j} - \xi \ox  \varrho^0 \lambda_0 - \xi \ox \varrho^j_{\rm a} \lambda_j  $.
Due to the affine structure, there is no way to modify the term $ - \xi \ox  \varrho^0 \lambda_0 $ which is the ``carrier'' of the 3-qubit entanglement by means of LOCC (or local filtering).
The result is the same for any ancilla, even entangled (even if we replace the 4-coherences term $ \xi \ox \varrho^j_{\rm a} \lambda_j  $ with something nontensorial).

\paragraph*{Violations of LHV models.}
The UPB state is known to be bound entangled but no Bell inequality is available for its detection.
In \cite{Cla-qu-ent-reflx1} we pointed out a class of multilinear algebraic inequalities which follow from the reflection symmetry and which lead to a simple form of contradiction of LHV models.
Briefly (all details are in \cite{Cla-qu-ent-reflx1}) such a contradiction has to do with the sign of the 2-coherences of the reflection of a separable density in $ {\cal C} $.
Unlike 1- and 3-coherences, such sign pattern cannot be reproduced by simply changing sign in the Bloch vectors of the original convex sum because of a parity condition.
If the original density has high rank and is close to the random state, then it is possible to compensate for the sign mismatch by adding a few more suitable terms in a mixture, but if the state is at the border of $ {\cal C} $ and has rank 4, like $ \rho_{\rm UPB} $, the existence of a convex combination representing all 3-coherences of $ \rho_{\rm UPB} $ and at the same time matching the ``odd'' sign pattern of its 2-coherences becomes unthinkable.
This to justify why the inequalities reported below are not fully fledged entanglement criteria but rather contradictions, indicating a violation of a LHV model.
Each of the 4 inequalities which can be obtained from (7b) of \cite{Cla-qu-ent-reflx1} is corresponding to one of the 4 components of $ \rho_{\rm sep} $ in \eqref{eq:rho_UPBsep}, and each unveils a sign pattern which is not compatible with a LHV model:
\beq
\begin{split}
1) \quad  \tr{ \rho_{\rm UPB} \Lambda_{031} } \tr{ \rho_{\rm UPB} \Lambda_{301} } \tr{ \rho_{\rm UPB} \Lambda_{330} } < 0 , \\
2) \quad  \tr{ \rho_{\rm UPB} \Lambda_{013} } \tr{ \rho_{\rm UPB} \Lambda_{103} } \tr{ \rho_{\rm UPB} \Lambda_{130} } < 0 , \\
3) \quad  \tr{ \rho_{\rm UPB} \Lambda_{033} } \tr{ \rho_{\rm UPB} \Lambda_{103} } \tr{ \rho_{\rm UPB} \Lambda_{130} } < 0 , \\
4) \quad  \tr{ \rho_{\rm UPB} \Lambda_{011} } \tr{ \rho_{\rm UPB} \Lambda_{101} } \tr{ \rho_{\rm UPB} \Lambda_{110} } < 0 .
\end{split}
\label{eq:ineq-LHV}
\eeq
Consider for example the case 1). 
Assuming the existence of a LHV model, the measure ${''+''}$ along $ \Lambda_{031} $ (we are only interested in its sign) must be compatible with the local measures $ f_j (\lambda_k) $ carried out on the $j$-th qubit along the $ \lambda_k  $ axis.
Two cases are possible: $ i) \;  f_2(\lambda_3 ) = {''+''} $, $  f_3(\lambda_1 ) = {''+''} $; $ii) \;  f_2(\lambda_3 ) = {''-''} $, $ f_3(\lambda_1 ) = {''-''} $.
Similarly, along $ \Lambda_{301} $ we measure ${''-''}$, hence $ i) \Rightarrow \;  f_1(\lambda_3 ) = {''-''} $; $ii) \Rightarrow \;  f_1 (\lambda_3 ) = {''+''} $.
Since $ \Lambda_{330} $ has measure ${''+''}$, both the LHV models $ i) $ and $ ii) $ are contradictory.
The inequalities \eqref{eq:ineq-LHV} are just a compact way to express these LHV violations. 
Replacing $ \rho_{\rm UPB} $ with $ \rho_{\rm sep} $, all 4 inequalities \eqref{eq:ineq-LHV} change sign and the contradictions disappear.
The argument used is of the same type of \cite{Mermin1}.
Since we have mixed states, we cannot formulate it rigorously in terms of ``signs of the eigenvalues'' of wavefunctions that are eigenfunctions as in the formulation of \cite{Mermin1,Peres2}.
However, assumption (5) of \cite{Cla-qu-ent-reflx1} is here verified and guarantees that each of the observables we consider (i.e., all those with nonzero expectation) can be traced back uniquely to one of the 4 ket states $ \ket{\psi_j} $.

\paragraph*{Preparation.}
The second question we are interested in is how to create the state $ \rho_{\rm UPB } $. 
While $ \rho_{\rm sep } $ is a mixture of separable pure states and can be produced by standard methods, the reflection operation $ \bar{S}_{64} $ is nonunitary and no quantum circuit is known for it.
It is obvious that if we start with any separable state and consider the orbit given infinitesimally by the Lie algebra $ {\rm Lie} \{ -i {\rm ad}_{\Lambda_{jkl }} \} $, $ j, k,l,= 0,1,2,3$, this will certainly contain all states (including entangled) and only states.
In particular, it is possible to find explicitly the Hamiltonian of a unitary transformation that maps $ \rho_{\rm sep } $ to $ \rho_{\rm UPB } $.
This is the following concatenation of nonlocal constant infinitesimal generators:
\beq
-i {\rm ad}_{H} = \begin{cases}
  -i {\rm ad}_{\Lambda_{333}} \qquad  \qquad  \qquad \; \; \text{for $ t\in [0, \, \frac{\tau_{p}}{2}]$} \\
 -i {\rm ad}_{\left( \Lambda_{011}+ \Lambda_{033}+\Lambda_{101}+\Lambda_{110}+ \Lambda_{303}+\Lambda_{330}\right)} \\
\qquad \qquad \qquad  \qquad   \qquad  \text{for $ t\in [ \frac{\tau_{p}}{2}, \,\frac{3 \tau_{p}}{4} ]$} 
\end{cases} 
\label{eq:Ham-creat1}
\eeq
% see file run_composite3_UPB, run_composite2_UPB
where $ \tau_p = 2 \sqrt{2} \pi $ is the period in the parametrization we are using.
The effect of the first part is to switch the sign of the entire homogeneous tensor $\xi $, except for the 6 2-coherences with pairwise equal indexes $ \varrho^{\{011,033,101,110,303,330\}} $.
In spite of the nonlocality of $ {\rm exp} \left( -it {\rm ad}_{\Lambda_{333}} \right) $, its action on $  \rho_{\rm sep} $ has remarkable ``local-like'' properties at the end of the time interval $ t= \frac{\tau_p}{2} $.
In fact, it changes sign to all and only the components of index ``1'' in each of the 3 qubits of the 4 components of $ \rho_{\rm sep}$.
This can be understood by means of the tensored Rodrigues' formula presented in \cite{Cla-spin-tens1}, which provides an explicit closed form expression for one-parameter flows such as each piece of \eqref{eq:Ham-creat1}.
For example:
\beqa
% \begin{split}
& {\rm exp} \left( -it {\rm ad}_{\Lambda_{333}} \right)  = \sum_{k=0} ^\infty \frac{(-i t )^k}{k!}  {\rm ad}_{\Lambda_{333}}^k = & \label{eq:exp-adL333} \\
&= I_4^{ \ox 3}  - i \sqrt{2}  \sin (\frac{t}{\sqrt{2}})  {\rm ad}_{\Lambda_{333} } - 2 \left(1 - \cos (\frac{t}{\sqrt{2}}) \right) {\rm ad}_{\Lambda_{333} } ^2 , & \nonumber
\eeqa
where $
 {\rm ad}_{\Lambda_{333}}^k  \! \! \! \! = \frac{1}{4^k} \!  (  
{\rm ad}_{\lambda_3} ^k  \! 
\ox {\rm aad}_{\lambda_3} ^k  \!
\ox {\rm aad}_{\lambda_3} ^k  \! \!
+   {\rm aad}_{\lambda_3} ^k  \!
\ox {\rm ad}_{\lambda_3} ^k  \! 
\ox {\rm aad}_{\lambda_3} ^k 
+   {\rm aad}_{\lambda_3} ^k 
\ox {\rm aad}_{\lambda_3} ^k 
\ox {\rm ad}_{\lambda_3} ^k 
+   {\rm ad}_{\lambda_3} ^k 
\ox {\rm ad}_{\lambda_3} ^k 
\ox {\rm ad}_{\lambda_3} ^k ) $,
with $ {\rm ad}_{\lambda_3}= \sqrt{2} i ( \delta_{32} - \delta_{23} ) $, $  {\rm ad}_{\lambda_3}^2 = 2  ( \delta_{22} + \delta_{33} ) $, $ {\rm aad}_{\lambda_3}= \sqrt{2}  ( \delta_{14} + \delta_{41} ) $, $  {\rm aad}_{\lambda_3}^2 = 2  ( \delta_{11} + \delta_{44} ) $.
Calling $ \rho_{\rm int} = {\rm exp} \left( -i  \frac{\tau_{p}}{2} {\rm ad}_{\Lambda_{333}} \right) \rho_{\rm sep} $, then we have $\rho_{\rm int} = \frac{1}{4} \sum_{j=1}^4  \ket{\mu_j} \bra{\mu_j}  $ with $ \ket{\mu_j}= \ket{01-},\ket{1-0},\ket{-01},\ket{+++}$.
%$ \rho_{\rm int} = \frac{1}{4} \sum_{j=1}^4  \rho_{{\rm int}_j}  $ with
% \beq
% \begin{split}
% \rho_{{\rm int}_1} & =  
%     \left( \varrho^0 \lambda_0 + \varrho^{3} \lambda_{3} \right) 
%\ox \left( \varrho^0 \lambda_0 - \varrho^{3} \lambda_{3} \right) 
%\ox \left( \varrho^0 \lambda_0 - \varrho^{1} \lambda_{1} \right) 
%\\
%\rho_{{\rm int}_2} & =  
%    \left( \varrho^0 \lambda_0 - \varrho^{3} \lambda_{3} \right) 
%\ox \left( \varrho^0 \lambda_0 - \varrho^{1} \lambda_{1} \right) 
%\ox \left( \varrho^0 \lambda_0 + \varrho^{3} \lambda_{3} \right) 
%\\
%\rho_{{\rm int}_3} & =  
%    \left( \varrho^0 \lambda_0 - \varrho^{1} \lambda_{1} \right) 
%\ox \left( \varrho^0 \lambda_0 + \varrho^{3} \lambda_{3} \right) 
%\ox \left( \varrho^0 \lambda_0 - \varrho^{3} \lambda_{3} \right) 
%\\
%\rho_{{\rm int}_4} & =  
%    \left( \varrho^0 \lambda_0 + \varrho^{1} \lambda_{1} \right) 
%\ox \left( \varrho^0 \lambda_0 + \varrho^{1} \lambda_{1} \right) 
%\ox \left( \varrho^0 \lambda_0 + \varrho^{1} \lambda_{1} \right) .
%\end{split}
%\label{eq:rho_UPBsep-int}
%\eeq 
Hence $ \rho_{\rm int} $ is still a separable state with the same properties as $ \rho_{\rm sep} $.
The second piece of \eqref{eq:Ham-creat1} is the one crucial for the purposes of creating the entanglement.
Its action can be analyzed by means of arguments similar to the ones above.
Both pieces of \eqref{eq:Ham-creat1} are nonlocal and they must be so because of the randomness of all  1-qubit densities.
While at the begin and end of each interval the density is PPT, during both evolutions there is always bipartite entanglement through each 1-2 cut.
Exchanging the order of application of the two Hamiltonians of \eqref{eq:Ham-creat1}, one still gets a map $ \rho_{\rm sep } \to \rho_{\rm UPB} $.
However the intermediate state reached is different: $  \tilde{\rho}_{\rm int} = {\rm exp} \left(  -i \frac{\tau_{p}}{4} {\rm ad}_{\left( \Lambda_{011}+ \Lambda_{033}+\Lambda_{101}+\Lambda_{110}+ \Lambda_{303}+\Lambda_{330}\right)} \right) \rho_{\rm sep } = \bar{S}_{64} ( \rho_{\rm int} )$.
It is straightforward to check that also $ \tilde{\rho}_{\rm int} $ obeys to \eqref{eq:ineq-LHV}, hence it is bound entangled.

\paragraph*{A PPT entangled orbit.}
Perturb $ \rho_{\rm UPB} $ by applying a Hamiltonian for a certain time interval.
%, choosing the Hamiltonian to be any of the basis directions $ -i {\rm ad}_{\Lambda_{jkl}} $, $ j, k,l \in \{ 0, 1,2,3\} $.
From randomness of all the 1-qubit reduced densities, we have that $ \rho_{\rm UPB} $ is (trivially) invariant to all local actions, as already mentioned above. 
When a 2- or 3- spin Hamiltonian is applied, the resulting density is generically not PPT.
There are at least two exceptions: the directions $ -i {\rm ad}_{ \left( \Lambda_{011}+ \Lambda_{022}+ \Lambda_{033}+\Lambda_{101}+\Lambda_{110}+ \Lambda_{202}+\Lambda_{220}+ \Lambda_{303}+\Lambda_{330}\right) } $ and $ -i  {\rm ad}_{\Lambda_{222}} $.
While the first generator induces no action at all on $ \rho_{\rm UPB} $ (i.e., $ \rho_{\rm UPB} $ is a fixed point for it), along the one-parameter orbit of the second one the density matrix is always PPT with respect to any bipartite cut.
Each density operator on such orbits has rank 4.
This orbit does not modifies any of the 6 1- or 2-qubit reduced densities, but affects only the $8$ 3-qubit coherences: $ \varrho^{111}$, $ \varrho^{113}$, $ \varrho^{131}$, $ \varrho^{133}$, $ \varrho^{311}$, $ \varrho^{313}$, $ \varrho^{331}$, $ \varrho^{333}$, i.e., exactly those 3-coherences that certainly do not yield bipartite entanglement (no index ${''2''}$, see \cite{Cla-qu-ent1}).
We can look at what happens for $ \rho_{\rm orb} (t) = {\rm exp} \left( -i t {\rm ad}_{\Lambda_{222}} \right) \rho_{\rm sep} $ and then consider the corresponding reflection $ \tilde{\rho}_{\rm orb} (t) = \bar{S}_{64} \left( \rho_{\rm orb} (t) \right)$ \footnote{In this case, in fact, the reflection commutes with the flow: $  \bar{S}_{64} \left(  {\rm exp} \left( -i t {\rm ad}_{\Lambda_{222}} \right) \rho_{\rm sep}  \right) =  {\rm exp} \left( -i t {\rm ad}_{\Lambda_{222}} \right) \bar{S}_{64} \left(  \rho_{\rm sep}  \right) $.}. 
Computing the sum of the series as in \eqref{eq:exp-adL333}, 
\begin{subequations}
\label{eq:exp-adL222-all}
\beqa
\rho_{\rm orb} (t) & 
= & \rho_{\rm sep} - \sin ( \frac{t}{\sqrt{2} } ) x \Lambda_{\{113,131,311,333\} }\nonumber  \\  && 
+ \left( 1 - \cos ( \frac{t}{\sqrt{2} } ) \right)  x \Lambda_{\{111,133,313,331\} }  \label{eq:exp-adL222-rhosep-a}\\
& = & \varrho^{\{ \alpha \} } \Lambda_{\{ \alpha \} }  
- \sin ( \frac{t}{\sqrt{2} } ) x \Lambda_{\{113,131,311,333\} } 
\nonumber \\ & &
- \cos ( \frac{t}{\sqrt{2} } ) x \Lambda_{\{111,133,313,331\} } ,
\label{eq:exp-adL222-rhosep-b}
\eeqa
\end{subequations}
where $ \{ \alpha \} $ denotes the subset of indexes $ jkl $ containing all 0- and 2-coherences of $ \rho_{\rm sep} $: $   \alpha  =  000,011,013,031,033,101,103,110,130,301,303,310,330 $.
In words, on the one-parameter orbit \eqref{eq:exp-adL222-all} all the 2-coherences are constants of the motion, while the 3-coherences evolve according to a sinusoidal law.
Neither of the last two terms alone in \eqref{eq:exp-adL222-rhosep-b} is a density (they are both traceless).
The sinusoidal law in \eqref{eq:exp-adL222-all} has the effect of ``swapping'' the bound entanglement between $ \tilde{\rho}_{\rm orb } $ and $ \rho_{\rm orb } $.
In fact, if $ \rho_{\rm orb }(0) = \rho_{\rm sep }  $ and  $ \tilde{\rho}_{\rm orb }(0) = \rho_{\rm UPB } $, at $ t=\frac{\tau_p}{4} $, $ \sin ( \frac{\tau_p}{4 \sqrt{2}} ) = 1 $ and we have to replace the 4 3-coherences of \eqref{eq:varrho-UPB} and \eqref{eq:rho_UPBsep} with the other 4. Denoting $ \rho_{\rm oq}= \rho_{\rm orb} ( \frac{\tau_p}{4 } )$: 
\beq
\rho_{\rm oq}:
\begin{cases}
& \varrho^{000} = \frac{1}{2\sqrt{2} } \\
& \varrho^{\{ 011,013,101,110,130,301 \} }=x  \\
& \varrho^{\{031,033,103,113,131,303,310,311,330,333 \}}=-x  \\
& \varrho^{jkl} =0 \qquad \text{otherwise}. \\
\end{cases}
\label{eq:rho-orb-tp/4}
\eeq 
It is easy to check that in this case $ \tilde{\rho}_{\rm orb }  ( \frac{\tau_p}{4 } ) $ is separable and given by $  \tilde{\rho}_{\rm orb }  ( \frac{\tau_p}{4 } ) = \frac{1}{4} \sum_{j=1}^4 \tilde{\rho}_{{\rm oq }_j} $:
\beq
\begin{split}
 \tilde{\rho}_{{\rm oq}_1} & =  
    \left( \varrho^0 \lambda_0 + \varrho^{1} \lambda_{1} \right) 
\ox \left( \varrho^0 \lambda_0 - \varrho^{3} \lambda_{3} \right) 
\ox \left( \varrho^0 \lambda_0 - \varrho^{1} \lambda_{1} \right) ,
\\
 \tilde{\rho}_{{\rm oq}_2} & =  
    \left( \varrho^0 \lambda_0 - \varrho^{1} \lambda_{1} \right) 
\ox \left( \varrho^0 \lambda_0 + \varrho^{1} \lambda_{1} \right) 
\ox \left( \varrho^0 \lambda_0 - \varrho^{3} \lambda_{3} \right) ,
\\
 \tilde{\rho}_{{\rm oq}_3} & =  
    \left( \varrho^0 \lambda_0 - \varrho^{3} \lambda_{3} \right) 
\ox \left( \varrho^0 \lambda_0 - \varrho^{1} \lambda_{1} \right) 
\ox \left( \varrho^0 \lambda_0 + \varrho^{1} \lambda_{1} \right) ,
\\
 \tilde{\rho}_{{\rm oq}_4} & =  
    \left( \varrho^0 \lambda_0 + \varrho^{3} \lambda_{3} \right) 
\ox \left( \varrho^0 \lambda_0 + \varrho^{3} \lambda_{3} \right) 
\ox \left( \varrho^0 \lambda_0 + \varrho^{3} \lambda_{3} \right) , 
\end{split}
\label{eq:tilde-rho-orb-tp/4}
\eeq 
or, in terms of kets, $ \tilde{\rho}_{{\rm oq}_i} = \ket{\theta_j}\bra{\theta_j } $ with $ \ket{\theta_j } = \ket{+1-}, \ket{-+1}, \ket{1-+},\ket{000}$, which is another UPB basis.
Hence $ \rho_{\rm oq} $ (obtainable from \eqref{eq:mapUPB}) is bound entangled.
The explicitly known mixture of the separable state $  \tilde{\rho}_{\rm orb} (\frac{\tau_p}{4 } ) $ tells us that the triplets of observables to be used to detect violations of LHV similar to \eqref{eq:ineq-LHV} in $ \rho_{\rm oq} $ are now different:
\beq
\begin{split}
\tr{ \rho_{\rm oq}\Lambda_{031}} \tr{ \rho_{\rm oq}\Lambda_{101}} \tr{ \rho_{\rm oq}\Lambda_{130}}<0 , \\
\tr{ \rho_{\rm oq}\Lambda_{013}} \tr{ \rho_{\rm oq}\Lambda_{103}} \tr{ \rho_{\rm oq}\Lambda_{110}}<0 , \\
\tr{ \rho_{\rm oq}\Lambda_{011}} \tr{ \rho_{\rm oq}\Lambda_{301}} \tr{ \rho_{\rm oq}\Lambda_{310} }<0, \\
\tr{ \rho_{\rm oq}\Lambda_{033}} \tr{ \rho_{\rm oq}\Lambda_{303}} \tr{ \rho_{\rm oq} \Lambda_{330}}<0 .
\end{split}
\label{eq:triples-obs-tp/4}
\eeq
The explicit knowledge of the mixture of pure states is crucial in setting the inequalities \eqref{eq:ineq-LHV} and \eqref{eq:triples-obs-tp/4}.
Notice in fact that $ \rho_{\rm UPB} $ is compatible with the LHV model given by the triplets of \eqref{eq:triples-obs-tp/4} and, similarly, that $ \rho_{\rm orb} (\frac{\tau_p}{4 } ) $ is compatible with those of \eqref{eq:ineq-LHV}.

\begin{figure}[h]
\begin{center}
 \includegraphics[width=5cm]{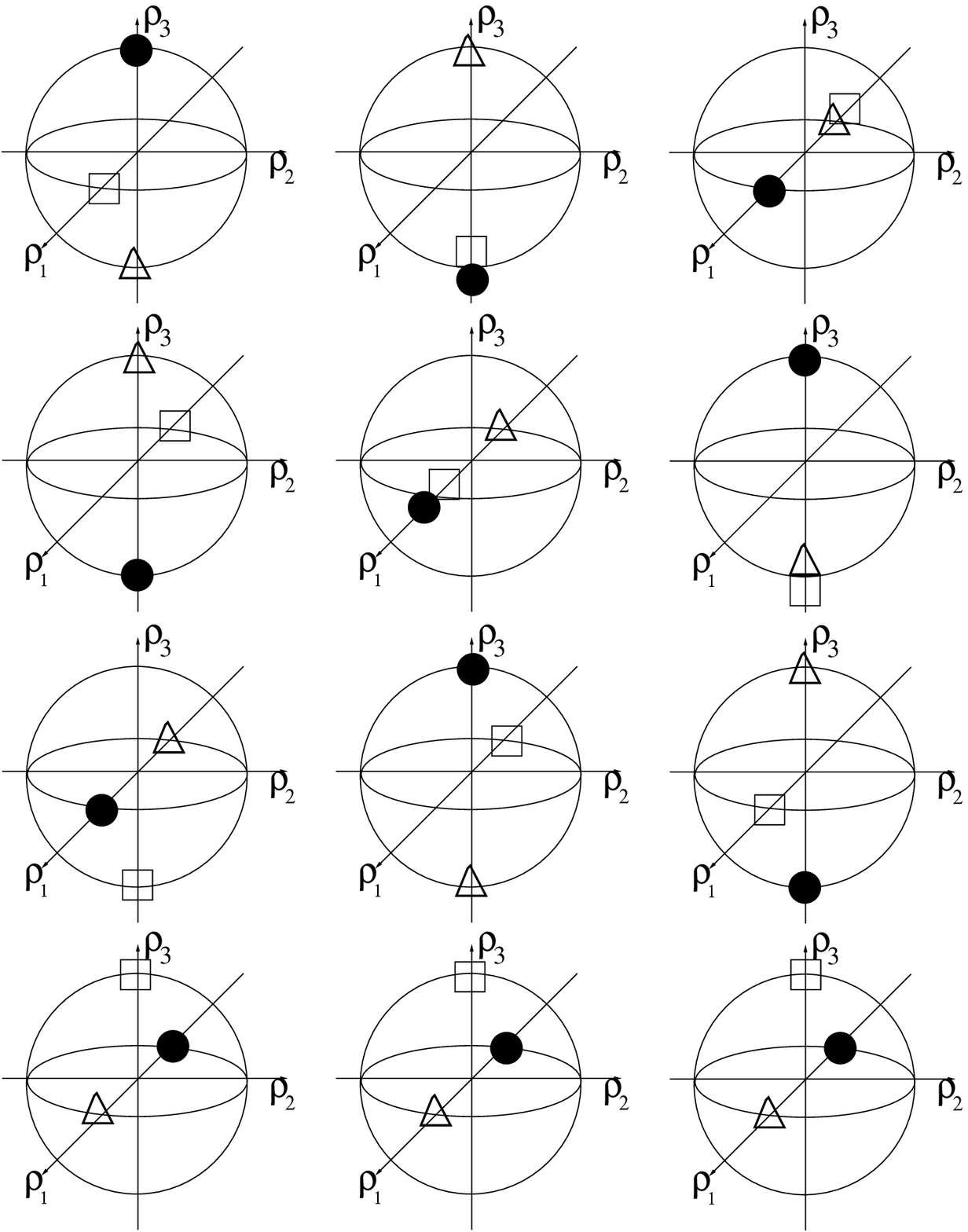}
 \caption{The Bloch vectors of the 4 pure states of the separable densities $ \rho_{\rm sep}=\rho_{\rm orb}(0 ) $ ($\bullet$), $ \tilde{\rho}_{\rm orb}(\frac{\tau_p}{4} ) $($\Box$) and $ \rho_{\rm orb}(\frac{\tau_p}{2} ) $ ($\triangle$).}
\label{fig:sp-3qubit-UPB1}
\end{center}
\end{figure}

After another fourth of period, the orbit $ \rho_{\rm orb}(t) $ reaches another separable state $  \rho_{\rm orb }(\frac{\tau_p}{2 } ) = \frac{1}{4} \sum_{j=1}^4 \ket{\phi_j} \bra{\phi_j } $, with $ \ket{\phi_j} = \ket{10-},  \ket{0-1}, \ket{-10}, \ket{+++} $, or, in terms of Bloch vectors, $  \rho_{\rm orb }(\frac{\tau_p}{2 } ) = \frac{1}{4} \sum_{j=1}^4 \rho_{{\rm oh }_j} $ with 
\[
\begin{split}
\rho_{{\rm oh}_1} & =  
    \left( \varrho^0 \lambda_0 - \varrho^{3} \lambda_{3} \right) 
\ox \left( \varrho^0 \lambda_0 + \varrho^{3} \lambda_{3} \right) 
\ox \left( \varrho^0 \lambda_0 - \varrho^{1} \lambda_{1} \right) ,
\\
\rho_{{\rm oh}_2} & =  
    \left( \varrho^0 \lambda_0 + \varrho^{3} \lambda_{3} \right) 
\ox \left( \varrho^0 \lambda_0 - \varrho^{1} \lambda_{1} \right) 
\ox \left( \varrho^0 \lambda_0 - \varrho^{3} \lambda_{3} \right) ,
\\
\rho_{{\rm oh}_3} & =  
    \left( \varrho^0 \lambda_0 - \varrho^{1} \lambda_{1} \right) 
\ox \left( \varrho^0 \lambda_0 - \varrho^{3} \lambda_{3} \right) 
\ox \left( \varrho^0 \lambda_0 + \varrho^{3} \lambda_{3} \right) ,
\\
\rho_{{\rm oh}_4} & =  
    \left( \varrho^0 \lambda_0 + \varrho^{1} \lambda_{1} \right) 
\ox \left( \varrho^0 \lambda_0 + \varrho^{1} \lambda_{1} \right) 
\ox \left( \varrho^0 \lambda_0 + \varrho^{1} \lambda_{1} \right)  .
\end{split}
%\label{eq:rho-orb-tp/2}
\]
Since $ \cos( \frac{\tau_p}{2\sqrt{2} } ) = -1 $, $ \tilde{\rho}_{\rm orb }  ( \frac{\tau_p}{4 } ) $ has the same tensor \eqref{eq:varrho-UPB} as $ \rho_{\rm UPB}$, but with the opposite sign in the 3-coherences. 
Comparing $\rho_{\rm sep}$ and $  \rho_{\rm orb }(\frac{\tau_p}{2 } ) $ (see Fig.~\ref{fig:sp-3qubit-UPB1}), $ \rho_{\rm orb }(\frac{\tau_p}{2 } ) $ corresponds to a change of the original UPB basis in which each member of the ket $ \ket{\phi_j}  $ is obtained by means of a $\pi $ rotation on the corresponding Bloch sphere around the $ \lambda_2 $ axis.
For example, $ \ket{\psi_1} = \ket{0} \ox \ket{1} \ox \ket{+} $ rotated around $ \Lambda_{222} = \lambda_2 \ox \lambda_2 \ox \lambda_2 $ gives $ \ket{\phi_1} = \ket{1} \ox \ket{0} \ox \ket{-} $ (first row in Fig.~\ref{fig:sp-3qubit-UPB1}) and so on.
A similar relation holds between $ \tilde{\rho}_{\rm orb }  ( \frac{\tau_p}{4 } ) $ and $ \tilde{\rho}_{\rm orb }  ( \frac{3 \tau_p}{4 } ) $.
The effect of the nonlocality of the rotation applied is the swapping of bound entanglement between  $ \tilde{\rho}_{\rm orb } $ and $\rho_{\rm orb } $, plus the recombination of pieces of product states into different (rotated) product states.
While the ``blocking'' state $ \rho_{{\rm sep}_4} $ undergoes a rotation around $ \lambda_2 $ on each qubit (plus ``swapping'') to yield $ \tilde{\rho}_{{\rm oq}_4} $ at $ t=\frac{\tau_p}{4} $ and $ \rho_{{\rm oh}_4} $ at $ t=\frac{\tau_p}{2} $ (see last row of Fig.~\ref{fig:sp-3qubit-UPB1}), every other triplet of Bloch vectors of $ \tilde{\rho}_{{\rm oq}_j} $ comes from a $ \frac{\pi}{2} $ rotation of 3 Bloch vectors of $ \rho_{\rm sep}$ belonging to 3 different pieces $ \rho_{{\rm sep}_j} $, $ j=1,2,3$.
For example, $ \tilde{\rho}_{{\rm oq}_1} = \ket{+1-}\bra{+1-} $ (i.e., the squares in the first row of Fig.~\ref{fig:sp-3qubit-UPB1}) contains the $ \frac{\pi}{2} $ rotations of the first Bloch vector of $ \rho_{{\rm sep}_1} $, the second of $ \rho_{{\rm sep}_2} $ and the third of $ \rho_{{\rm sep}_3} $ (i.e., the $ \frac{\pi}{2} $ rotations of the diagonal bullets in Fig.~\ref{fig:sp-3qubit-UPB1}).
While at $ t= \frac{k \tau_p}{4} $, $ k \in \mathbb{N}$, a pair state/reflected state admits a neat splitting into separable and bound entangled, the situation is more ambiguous when $ t \neq \frac{k \tau_p}{4} $.
From \eqref{eq:exp-adL222-all}, the conjecture is that both $  \rho_{\rm orb }  (t ) $ and $  \tilde{\rho}_{\rm orb }  (t ) $ share a percentage of separability and entanglement in a ``nonseparable'' way.
There may be other nontrivial PPT entangled orbits, but they are more difficult to find.

Finally notice that, as a byproduct, one gets a simpler scheme for the preparation of $ \rho_{\rm UPB} $ from a separable state: $ \rho_{\rm UPB} = {\rm exp} ( i t {\rm ad}_{\Lambda_{222}} ) \rho_{\rm orb} ( \frac{\tau_p}{4} )= {\rm exp} (- i t {\rm ad}_{\Lambda_{222}} ) \rho_{\rm orb} ( \frac{3 \tau_p}{4} )$.

\bibliographystyle{apsrev}

% \bibliography{/home/altafini/tex/bib/quant}

\small

\begin{thebibliography}{12}
\expandafter\ifx\csname natexlab\endcsname\relax\def\natexlab#1{#1}\fi
\expandafter\ifx\csname bibnamefont\endcsname\relax
  \def\bibnamefont#1{#1}\fi
\expandafter\ifx\csname bibfnamefont\endcsname\relax
  \def\bibfnamefont#1{#1}\fi
\expandafter\ifx\csname citenamefont\endcsname\relax
  \def\citenamefont#1{#1}\fi
\expandafter\ifx\csname url\endcsname\relax
  \def\url#1{\texttt{#1}}\fi
\expandafter\ifx\csname urlprefix\endcsname\relax\def\urlprefix{URL }\fi
\providecommand{\bibinfo}[2]{#2}
\providecommand{\eprint}[2][]{\url{#2}}

\bibitem[{\citenamefont{Bennett et~al.}(1999)\citenamefont{Bennett, DiVincenzo,
  Mor, Shor, Smolin, and Terhal}}]{Bennett1}
\bibinfo{author}{\bibfnamefont{C.~H.} \bibnamefont{Bennett}},
  \bibinfo{author}{\bibfnamefont{D.~P.} \bibnamefont{DiVincenzo}},
  \bibinfo{author}{\bibfnamefont{T.}~\bibnamefont{Mor}},
  \bibinfo{author}{\bibfnamefont{P.~W.} \bibnamefont{Shor}},
  \bibinfo{author}{\bibfnamefont{J.~A.} \bibnamefont{Smolin}},
  \bibnamefont{and} \bibinfo{author}{\bibfnamefont{B.~M.}
  \bibnamefont{Terhal}}, \bibinfo{journal}{Phys. Rev. Lett.}
  \textbf{\bibinfo{volume}{82}}, \bibinfo{pages}{5385} (\bibinfo{year}{1999}).

\bibitem[{\citenamefont{Horodecki et~al.}(1998)\citenamefont{Horodecki,
  Horodecki, and Horodecki}}]{Horodecki7}
\bibinfo{author}{\bibfnamefont{M.}~\bibnamefont{Horodecki}},
  \bibinfo{author}{\bibfnamefont{P.}~\bibnamefont{Horodecki}},
  \bibnamefont{and}
  \bibinfo{author}{\bibfnamefont{R.}~\bibnamefont{Horodecki}},
  \bibinfo{journal}{Phys. Rev. Lett.} \textbf{\bibinfo{volume}{80}},
  \bibinfo{pages}{5239} (\bibinfo{year}{1998}).

\bibitem[{\citenamefont{Smolin}(2001)}]{Smolin1}
\bibinfo{author}{\bibfnamefont{J.~A.} \bibnamefont{Smolin}},
  \bibinfo{journal}{Phys. Rev. A} \textbf{\bibinfo{volume}{63}},
  \bibinfo{pages}{032306} (\bibinfo{year}{2001}).

\bibitem[{\citenamefont{D{\"u}r}(2001)}]{Dur1}
\bibinfo{author}{\bibfnamefont{W.}~\bibnamefont{D{\"u}r}},
  \bibinfo{journal}{Phys. Rev.Lett.} \textbf{\bibinfo{volume}{87}},
  \bibinfo{pages}{230402} (\bibinfo{year}{2001}).

\bibitem[{\citenamefont{DiVincenzo et~al.}(2003)\citenamefont{DiVincenzo, Mor,
  Shor, Smolin, and Terhal}}]{DiVincenzo1}
\bibinfo{author}{\bibfnamefont{D.~P.} \bibnamefont{DiVincenzo}},
  \bibinfo{author}{\bibfnamefont{T.}~\bibnamefont{Mor}},
  \bibinfo{author}{\bibfnamefont{P.~W.} \bibnamefont{Shor}},
  \bibinfo{author}{\bibfnamefont{J.~A.} \bibnamefont{Smolin}},
  \bibnamefont{and} \bibinfo{author}{\bibfnamefont{B.~M.}
  \bibnamefont{Terhal}}, \bibinfo{journal}{Comm. Math. Phys.}
  \textbf{\bibinfo{volume}{238}}, \bibinfo{pages}{379} (\bibinfo{year}{2003}).

\bibitem[{\citenamefont{Horodecki}(1997)}]{Horodecki2}
\bibinfo{author}{\bibfnamefont{P.}~\bibnamefont{Horodecki}},
  \bibinfo{journal}{Phys. Lett. A} \textbf{\bibinfo{volume}{232}},
  \bibinfo{pages}{333} (\bibinfo{year}{1997}).

\bibitem[{\citenamefont{Altafini}(2004{\natexlab{a}})}]{Cla-qu-ent-reflx1}
\bibinfo{author}{\bibfnamefont{C.}~\bibnamefont{Altafini}},
  \bibinfo{journal}{Preprint arXiv:quant-ph/0405123}
  (\bibinfo{year}{2004}{\natexlab{a}}).

\bibitem[{\citenamefont{Altafini}(2004{\natexlab{b}})}]{Cla-qu-ent1}
\bibinfo{author}{\bibfnamefont{C.}~\bibnamefont{Altafini}},
  \bibinfo{journal}{Physical Review A} \textbf{\bibinfo{volume}{69}},
  \bibinfo{pages}{012311} (\bibinfo{year}{2004}{\natexlab{b}}).

\bibitem[{\citenamefont{Terhal et~al.}(2003)\citenamefont{Terhal, Doherty, and
  Schwab}}]{Terhal2}
\bibinfo{author}{\bibfnamefont{B.~M.} \bibnamefont{Terhal}},
  \bibinfo{author}{\bibfnamefont{A.~C.} \bibnamefont{Doherty}},
  \bibnamefont{and} \bibinfo{author}{\bibfnamefont{D.}~\bibnamefont{Schwab}},
  \bibinfo{journal}{Phys. Rev. Lett.} \textbf{\bibinfo{volume}{91}},
  \bibinfo{pages}{157903} (\bibinfo{year}{2003}).

\bibitem[{\citenamefont{Mermin}(1994)}]{Mermin1}
\bibinfo{author}{\bibfnamefont{N.~D.} \bibnamefont{Mermin}},
  \bibinfo{journal}{Rev. Mod. Phys.} \textbf{\bibinfo{volume}{65}},
  \bibinfo{pages}{803} (\bibinfo{year}{1994}).

\bibitem[{\citenamefont{Peres}(1993)}]{Peres2}
\bibinfo{author}{\bibfnamefont{A.}~\bibnamefont{Peres}},
  \emph{\bibinfo{title}{Quantum theory : concepts and methods}}
  (\bibinfo{publisher}{Kluwer}, \bibinfo{year}{1993}).

\bibitem[{\citenamefont{Altafini}(2004{\natexlab{c}})}]{Cla-spin-tens1}
\bibinfo{author}{\bibfnamefont{C.}~\bibnamefont{Altafini}},
  \bibinfo{journal}{Preprint arXiv:quant-ph/0404006}
  (\bibinfo{year}{2004}{\natexlab{c}}).

\end{thebibliography}

\end{document}